\documentclass[twocolumn,showpacs,preprintnumbers,amsmath,amssymb,10pt,aps,prb]{revtex4-2}
\usepackage{epsfig,amsopn}
\usepackage{graphicx}
\usepackage{epstopdf}
\usepackage{amsmath,amssymb}
\usepackage{amsthm}
\usepackage{enumerate}
\usepackage{xcolor}
\usepackage{fixmath}
\usepackage[colorlinks=true,allcolors=blue]{hyperref}

\begin{document}
	\title{Plasmon in Nonsymmorphic Dirac semimetals }
	\author{Debasmita Giri}
	\affiliation{Department of Physics, Indian Institute of Technology Kanpur, Kanpur 208016, India}
	\author{Arijit Kundu}
	\affiliation{Department of Physics, Indian Institute of Technology Kanpur, Kanpur 208016, India}
	
	\begin{abstract}
We study the collective charge-density modes (plasmons) of two-dimensional nonsymmorphic Dirac semimetals, within the random-phase approximation (RPA) in presence of  Coulomb interaction. Without loss of generality, we consider a system in a two-dimensional square-lattice, based on the model originally predicted by Young and Kane [\href{https://doi.org/10.1103/PhysRevLett.115.126803}{Phys. Rev. Lett. \textbf{115}, 126803 (2015)}], where the  non-interacting band-structure consists of three band-touching points, near which the electronic states follow Dirac equations. Two of these Dirac nodes, at the momentum points $X_1$ and $X_2$, are anisotropic, i.e., disperse with different velocities in different directions, whereas the third Dirac point at $M$ is isotropic. Interestingly we find that the system of these three Dirac nodes hold a single low-energy plasmon mode, within its particle-hole gap, that disperses in isotropic manner, in the case when the nodes at $X_1$ and $X_2$ are related by symmetry, which we further show in a long-wavelength approximation. We also discuss the effects of possible perturbations that can give rise to anisotropic plasmon dispersions. Our results suggest, in similarity with graphene, plasmon modes of such non-symmorphic semimetals are highly tunable and hold promise for possible applications.
	\end{abstract}
	\maketitle
	\section{Introduction}
	
	\begin{figure}[t]
	\begin{center}
		\includegraphics[width=0.3\textwidth]{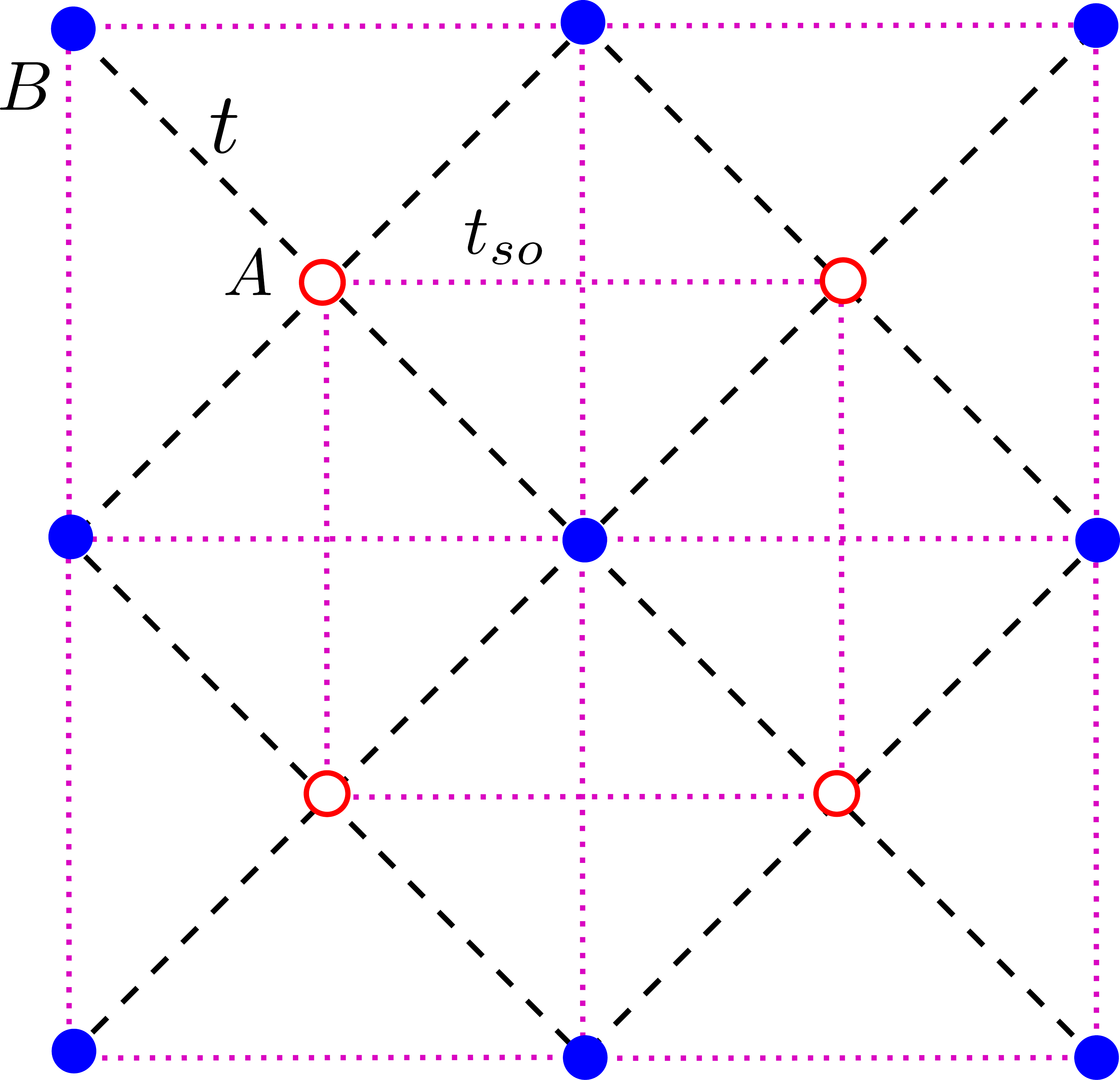}
		\caption{The schematic lattice model for a 2D nonsymmorphic Dirac semimetal, where each of the lattice-point consists of two sub-lattices $A$ and $B$. The $A$-$B$ hopping amplitude is $t$ and the spin-orbit interaction induced $A$-$A$ and $B$-$B$ hopping amplitude is $t_{so}$. }
		\label{lattice}
	\end{center}
\end{figure}
The two-dimensional (2D) Dirac systems have been a fascinating subject of research for the last decade, both theoretically and experimentally. These 2D Dirac systems, including graphene, possess several unique electronic and optical properties owing to their linear energy dispersion and nontrivial Berry phase.~\cite{2drev1,2drev2}.
 It has been a challenging task to find other two-dimensional materials with similar Dirac like dispersions and only a few such materials are known to host Dirac nodes, such as silicene~\cite{sil} and germanene~\cite{germ}, which are also difficult to find in a stable form.  The Dirac point in these materials is protected by symmetry in the absence of spin-orbit coupling (SOC), whereas finite SOC opens up gaps at the Dirac points. Recently, a new class of possible 2D Dirac semimetals has been introduced by Young and Kane \cite{Kane}, where the nonsymmorphic symmetry protects the Dirac points even in the presence of SOC~\cite{nons2, nons3, nons4, nons5}. Since then, there has been a surge in the search for such 2D materials with Dirac like band structure and several groups have reported possible candidates in recent times. Among the predicted stable 2D materials with nonsymmorphic space group symmetries which possess Dirac points in the presence of SOC, include group-VA 2D materials with phosphorene structure~\cite{VAP} and 2D materials with $\alpha$-SnO structure~\cite{SNO}, although these Dirac points are away from the Fermi level. 
  	Based on the first-principles calculations and theoretical analysis, Guan \textit{et al.}~\cite{HfGeTe} also proposed a family of stable 2D materials in the HfGeTe-family of monolayer systems, which host a pair of Dirac points close to the Fermi level in the presence of SOC. These Dirac points are called 2D spin-orbit Dirac points(SPDs)~\cite{SDP}. The 2D SPDs also appear in the other members of the HfGeTe-family, such as ML-HfSnTe and ML-HfSiTe, and materials with Te substituted by other chalcogens elements.  These materials resemble the 2D Dirac semimetal in the presence of SOC proposed by Young and Kane.
 
  	 The first experimental report of such 2D systems with Dirac nodes protected by nonsymmorphic symmetries was on $\alpha$-Bismuthene~\cite{bism}, which contains two anisotropic Dirac points, that matches closely with the original predictions by Young and Kane.  Other possible examples of such systems include predictions in 
  	 chemically modified group-VA (As, Sb and Bi) 2D puckered structure \cite{VA}.

In the presence of the electron-electron interaction, apart from changes in the band structure of the elementary excitation, there are also collective excitations. Plasmons are one of such collective modes that result from collective charge oscillations of the system~\cite{flensberg,giuliani}. The polarization function and plasmon modes have been studied extensively in 2D semimetals with Dirac like dispersion in the context of Graphene~\cite{dasg,dasg2,plg1,Wunsch}, surface of three-dimensional (3D) Weyl and Dirac semimetals ~\cite{weylplsurf1,weylplsurf2,weylplsurf3,weylplsurf4,weylplsurf5,weylplsurf6,weylplsurf7,weylplsurf8,weylplsurf9,weylplsurf10,weylplsurf11}, tilted Dirac semimetal~\cite{tild,amitan}, and the surface of 3D topological insulators~\cite{plti1,plti2,plti3,plti4}. Recently theory of anisotropic plasmon has also been investigated in Ref.~\onlinecite{dasan}.

In this work, we consider the effect of the Coulomb interaction in a nonsymmorphic Dirac semimetal giving rise to collective charge oscillation mode, especially focusing on the model originally predicted by Young and Kane~\cite{Kane} and possible realization in  $\alpha$-Bismuthene~\cite{bism}.
For the non-interacting system, we model the system in a two-dimensional square lattice, and we evaluate the density-density correlation function within the random-phase approximation (RPA). The non-interacting band structure consists of three band-touching points, near which the electronic states follow Dirac equations. Two of these Dirac nodes, at the momentum, points $X_1$ and $X_2$, are anisotropic, i.e., disperse with different Fermi velocities in different directions, whereas the third Dirac point at $M$ is isotropic. Interestingly we find that, the system of these three Dirac nodes holds a single low-energy plasmon mode, within its particle-hole gap, that disperses in an isotropic manner, as long as the dispersions of the Dirac nodes at $X_1$ and $X_2$ are related by $k_x \rightarrow k_y$ symmetry. Further, we show this analytically using a long-wavelength approximation. Finally, we comment on the case when such symmetry may not be present as well as when the system is perturbed.
	\begin{figure}[t]
	\begin{center}
		\includegraphics[width=0.45\textwidth]{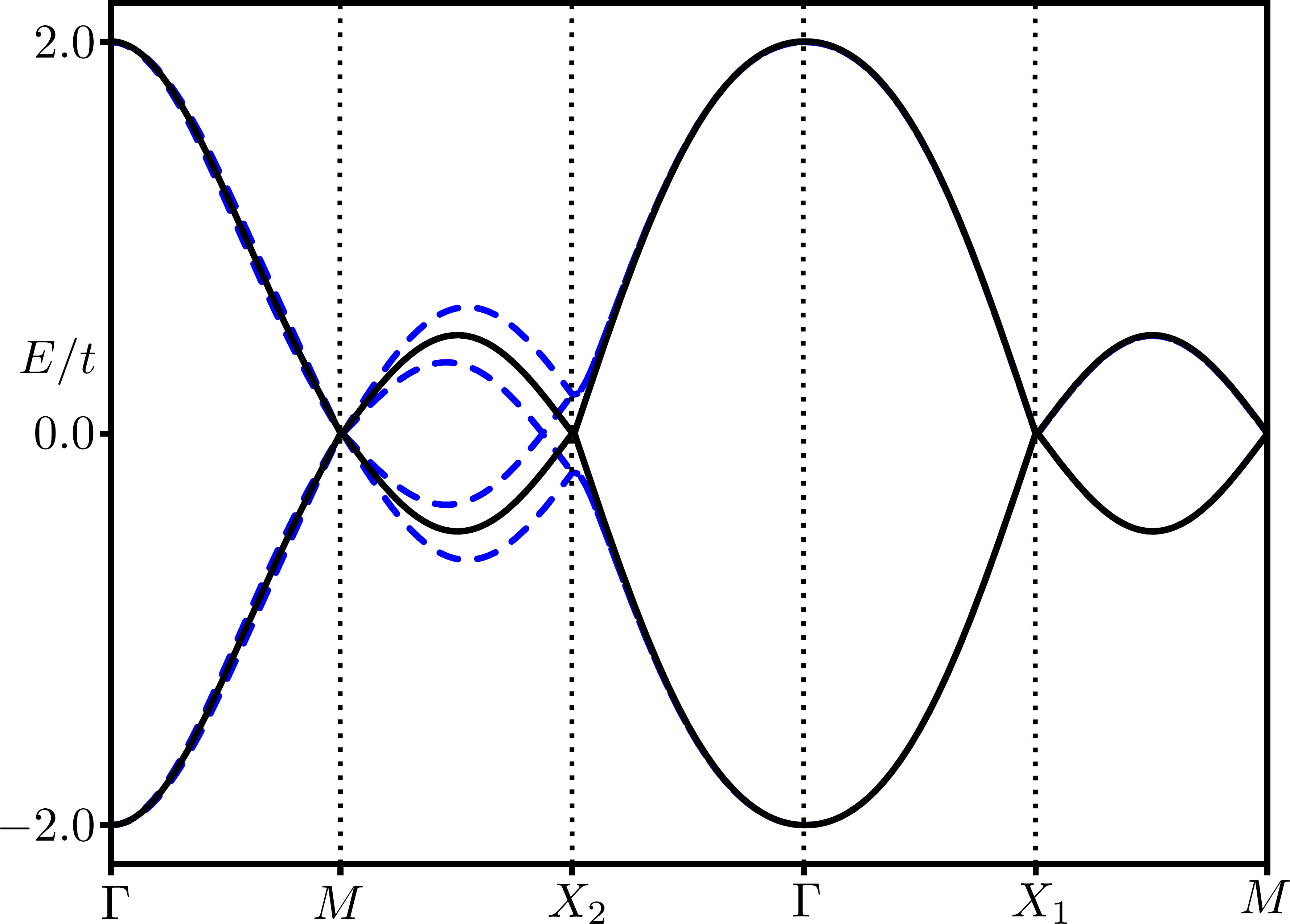}
		\caption{Solid lines represent the band-structure of the Hamiltonian Eq.~\ref{nonsym_tbm0} along a set of high-symmetry points, showing three band-touching points at $M$, $X_1$ and $X_2$ momentum. We used $t=1, t_{so}=0.5$ and $t_2=0$. The dashed lines represent the modified bands, in presence of the perturbation Eq.~\ref{eq:pert}, which splits the Dirac node at the $X_2$ into two Weyl nodes., and, modifies the dispersion of the Dirac node at the $M$ point.}
		\label{bands}
	\end{center}
\end{figure}

The paper is organized as following. We briefly introduce the nonsymmorphic systems in Sec. \ref{sec:model}. Theoretical background of dynamical polarization functions and the plasmon modes for the systems has been discussed in Sec. \ref{sec:dynpol}. In Sec. ~\ref{sec:result}, we present the numerical results from the tight-binding model as well as a long-wavelength approximation to compare with such numerical results. Finally, we discuss the case of anisotropic plasmon and the effect of perturbations in Sec.~\ref{sec:sum}.

\begin{figure*}[t]
\begin{center}
			\includegraphics[width=0.85\textwidth]{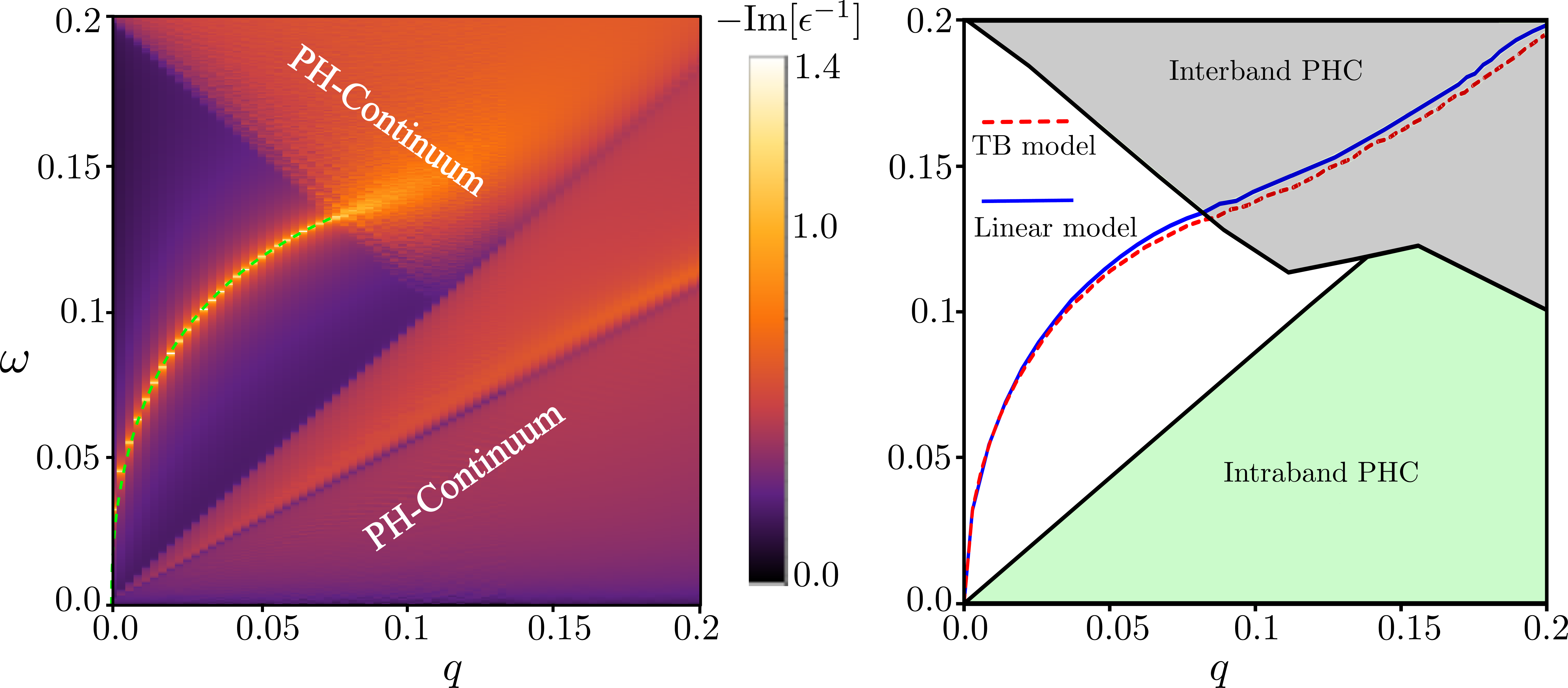}
			\caption{Left panel: Density plot of lorgarithm of the loss function, defined as Eq.~\ref{lossdefi}. The continuous regions where Im$[\epsilon^{-1}]\neq0$ represents the particle-hole continuum (PHC). The sharp bright line outside the PHC indicates the plasmon mode. Right panel: Plasmon dispersion along with PHC for $\mu=0.1$ and $\theta={\rm tan^{-1}} \frac{q_y}{q_x}$~=$\frac{\pi}{4}$. The dashed and the solid curves represent the plasmon dispersion using the tight-binding model Eq.~\ref{nonsym_tbm} and  the linearized Hamiltonian Eq.~\ref{linH}, respectively. The parameters we used are $t=1, t_{so}=0.5$, $t_2=0$, fine-structure constant $\alpha=1$ and chemical potential $\mu=0.1$.}
			\label{pl_dis}
\end{center}
\end{figure*}

	\section{Model}\label{sec:model}
	We begin with the tight-binding Hamiltonian, originally introduced by Young and Kane~\cite{Kane} for the nonsymmorphic Dirac semimetals on a square-lattice, Fig.~\ref{lattice}, with two atoms per unit cell; where one of the atoms is shifted out of the plane along the $\hat{z}$ direction, given by,
	\begin{align} 
     \tilde{H}_0=&2t\tau_x \cos\frac{k_x}{2}\cos\frac{k_y}{2} + t_{so}\tau_z\left(\sigma_y \sin k_x  - \sigma_x \sin k_y\right)\nonumber\\
&+ t_2 (\cos k_x + \cos k_y)	\label{nonsym_tbm0}
	\end{align}
$\sigma$ and $\tau$ are the Pauli matrices for spin and sublattice degrees of freedom; $t$ is the amplitude of nearest-neighbor hopping, and $t_{so}$ is the amplitude of the next-nearest-neighbor spin-orbit (SO) interaction. The momentum is made unit-less, by the multiplication of the lattice spacing $a_0$. The Brillouin zone is thus defined by $k_x \in \{-\pi,\pi\}$ and $k_y \in \{-\pi,\pi\}$. In the band-structure of the above Hamiltonian, there are three band-touching Dirac points, situated at $X_1=\{\pi,0\}$, $X_2=\{0,\pi\}$ and $M=\{\pi,\pi\}$. Near these Dirac nodes, the low-energy electronic states follow the Dirac equations. The Hamiltonian above is symmetric under the usual parity and time-reversal operations. We neglected the next-nearest neighbor hopping term $t_2$, which simply shifts the energies of the Dirac nodes at $X_1, X_2$, compared to the Dirac node at $M$ point. This does not alter the main conclusions of the work, as we argued later. 

The existence and robustness of these Dirac nodes are protected by a nonsymmorphic symmetry as following~\cite{nonsym}. The Hamiltonian is invariant under three nonsymmorphic symmetries: (i) $g_1 = i\sigma_x\tau_x \hat{S}_x \hat{t}_{\frac12\hat{x}}$,  (ii) $g_2 = i\sigma_y\tau_x \hat{S}_y \hat{t}_{\frac12\hat{y}}$, and, (iii) $g_3 = i\sigma_z\tau_x t_{\frac12\hat{x}}t_{\frac12\hat{y}}$, where $\hat{t}_{\frac12 \hat{i}}$ denoted half-unite vector translation along the direction $\hat{i}$ and $\hat{S}_i$ denotes rotation by an angle $\pi$ around the $i$ axes. Presence of these symmetries implies that the Hamiltonian and these operators can be diagonalized simultaneously. Along a invariant line in the Brillouin-zone, where $g_i\vec{k}=\vec{k}$, the Bloch states, for some $n^{\rm th}$ band, can be chosen as eigenstates, $g_i |\psi_{n}^{\pm}(\vec{k})\rangle = \pm \lambda e^{i\vec{k}\cdot\hat{t}_{\frac12\hat{i}}}|\psi_{n}^{\pm}(\vec{k})\rangle$. As, $e^{i\vec{G}\cdot \hat{t}_{\frac12\hat{i}}}=-1$, these eigen-vectors must switch as one goes from a momentum $\vec{k}$ to another equivalent momentum $\vec{k}+\vec{G}$ along the $g_i$ invariant line. In addition, in the presence of the time-reversal as well as inversion symmetry, there must be a four-fold degenerate crossing at the time-reversal invariant point $\vec{k} = \vec{G}/2$. These points are Dirac points as the four-fold degeneracy splits away from these points~\cite{Kane,Param}. The $g_i$ invariant lines in the present system are following: $g_1$ invariant lines $k_y = 0,\pm \pi$; $g_2$ invariant lines $k_x = 0,\pm \pi$. The Dirac points are at the intersections of these invariant lines, $X_1 = \{\pi,0\}$, $X_2 = \{ 0, \pi\}$ and $M = \{\pi,\pi\}$. The band structure of the system is briefly sketched in Fig.~\ref{bands}.
  
For simplicity, we make a unitary transformation, such that the Hamiltonian becomes block-diagonal in the orbital basis, given by the unitary matrix~\cite{nonsym}
\begin{align}\label{eq:U}
U = \frac{1}{\sqrt{2}}\left(\begin{array}{cc} \sigma_0 & \sigma_x \\ \sigma_z & -i\sigma_y \end{array}\right),
\end{align}
to have, in the rotated basis,
	\begin{align} 
     H_0=d_1\sigma_x+d_2\tau_z \sigma_y +d_3\sigma_z .
	\label{nonsym_tbm}
	\end{align}
with,
	\begin{align}
		&d_1=-t_{so}\sin{k_y}, \nonumber \\
        &d_2=t_{so}\sin{k_x}, \nonumber \\
        &d_3=2t\cos\frac{k_x}{2}\cos\frac{k_y}{2} .		
	\end{align}
To proceed, we re-write this non-interacting Hamiltonian in the second-quantized notation, as
\begin{align*}
H_0=\sum \xi_{m,\vec{k}}~ c^\dagger_{m,\vec{k}} c_{m,\vec{k}}
\end{align*}
where $c_{m,\vec{k}}$ is the anhilation operator at momentum $k$ of $m^{th}$ band. In addition to the non-interacting Hamiltonian, we consider the electrons to be interacting through Coulomb interaction 
	\begin{align}\label{eq:Hint1}
	H_{\text{int}} = {1 \over 2}  \int d^3\vec{r}_1 d^3\vec{r}_2 \rho(\vec{r}_1)V(|\vec{r}_1-\vec{r}_2|) \rho(\vec{r}_2),
	\end{align}
 Here the charge density operator  is defined as $ \rho (\vec{r})= \Psi^{\dagger} (\vec{r}) \Psi (\vec{r})$
 and the field operator is, 
	$	\Psi(\vec{r})=\frac{1}{\sqrt{A}}\sum_{\vec{k}} \exp(i \vec{k}.\vec{r}) \sum_{m} \phi_{m,\vec{k}} ~ c_{m,\vec{k}}$, where $A$ is the area of the system. $\phi_{m.\vec{k}}$ are the spinors corresponding to the eigenstates $ |m,\vec{k}\rangle$.

 \section{Dynamical polarization function}\label{sec:dynpol}
 Plasmon modes are collective charge oscillation modes that move in the self-consistent field arising from the electrons interacting through the Coulomb interaction. These modes appear as the poles of the interacting \textit{density-density response function}, given by
	\begin{align}\label{eq:chi}
	\chi(\vec{r},\vec{r'}, t)= -\textit i  \Theta(t)\big \langle \big[\rho(\vec{r},t), \rho(\vec{r'})\big] \big\rangle.
	\end{align}
 where $\rho(\vec{r},t)= \exp(i  H t) \rho (\vec{r}) \exp(-i  H t)$ with $H=H_0+H_{\rm int}$.
If we write the the Coulomb interactions, Eq.~(\ref{eq:Hint1}), in the second-quantized notation as
	\begin{align}\nonumber
	H_{\text{int}} = {1 \over 2A}  \sum V(|\vec{q}_1|) \phi^\dagger_{l_1,\vec{k}_1}\phi_{l_4,\vec{k}_1-\vec{q}_1}\phi^\dagger_{l_2,\vec{k}_2}\phi_{l_3,\vec{k}_2+\vec{q}_1} \\ \times
	c^\dagger_{l_1,\vec{k}_1} 
	c^\dagger_{l_2,\vec{k}_2}c_{l_3,\vec{k}_2+\vec{q}_1}c_{l_4,\vec{k}_1-\vec{q}_1},\label{eq:Hint2}
	\end{align}
then the density response function in the Fourier space can be written as,
		\begin{align}\label{eq:chiq}
	\chi(\vec{q},\omega)&= -i\frac{\Theta(t)}{A}\int dt e^{i\omega t}\big\langle [\rho(\vec{q},t) , \rho(-\vec{q},0)] \big \rangle
	\end{align}
where $\Theta(\cdot)$ is the unit step function and 
\begin{align}\label{eq:rhoq}
  \rho(\vec{q})= \sum_{m,m^{\prime},\vec{k}} \phi^{\dagger}_{m,\vec{k}} \phi_{m^{\prime},\vec{k}+\vec{q}~}~ c^{\dagger}_{m,\vec{k}} c_{m^{\prime},\vec{k}+\vec{q}}~.
\end{align}
  
\begin{figure*}[htbp]
\begin{center}\leavevmode
 		\includegraphics[width=1.0\textwidth]{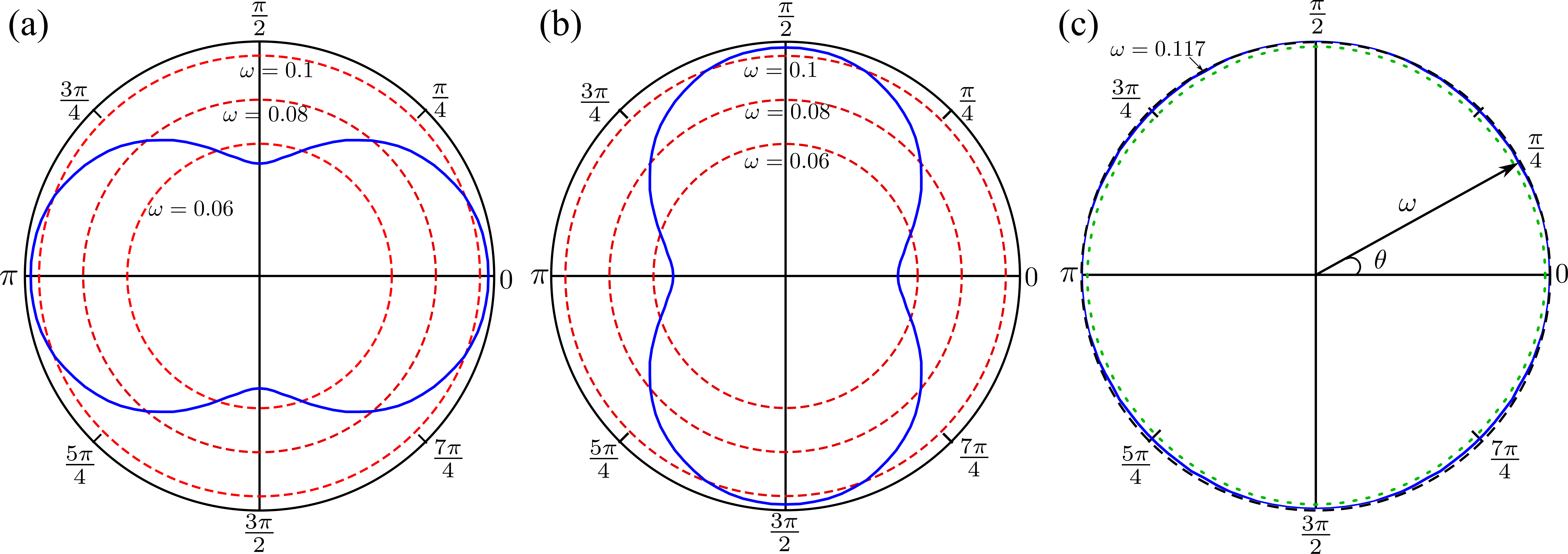}
 		\caption{ The blue solid curves represent the dispersions of the plasmon modes, for the system of Dirac nodes, Eq.~(\ref{linH}), as a function of $(\omega,\theta)$ for a fixed value of $q=0.05$, where $\theta={\rm tan^{-1}} \frac{q_y}{q_x}$. The red-dashed circles are the locus of constant $\omega$. Panel (a): the plasmon dispersion, if one considers a single Dirac node at $X_1=\{\pi,0\}$ in the band-structure. The dispersion clearly shows the anisotropic behavior of plasmon mode, i.e. $\omega$ changes as a function of $\theta$. Panel  (b): similarly, the plasmon dispersion, if one considers a single Dirac node at $X_2=\{0,\pi\}$  in the band-structure, which is also anisotropic in nature. Panel  (c): the plasmon dispersion of the nonsymmorphic Dirac semimetal which hosts three Dirac nodes, at momentum $X_1$, $X_2$ and $M$ in the band-structure.The dotted (green) contour of the plasmon dispersion is obtained from the tight-binding Hamiltonian, Eq.~(\ref{nonsym_tbm}). Although, if one considers individually the Dirac nodes $X_1$ and $X_2$, the plasmon dispersion is anisotropic, but the full system shows isotropic plasmon dispersion i.e, $\omega$ remains constant for $\theta=(0,2\pi)$. The parameters used are the same as in the Fig.~\ref{pl_dis}.  }
 		\label{iso}
\end{center}
\end{figure*}

In the linear response regime, within the random phase approximation (RPA) the interacting density-density response function is approximated by the following expression (see Appendix \ref{2nd_not} for details of the derivation),
	\begin{align}
	\chi(\vec{q},\omega)=\frac{\chi_0(\vec{q},\omega)}{1-V(|\vec{q}|)\chi_0(\vec{q},\omega)}.\label{eq:chiint}
	\end{align}

The plasmon modes are then obtained by solving the poles of the interacting density-density response function or, equivalently, by looking for zeros of the dielectric function,
\begin{align}
	\epsilon(\vec{q},\omega)=1-V(|\vec{q}|)\chi_0(\vec{q},\omega),\label{eq:epsilon}
\end{align}	
where, $V(\vec{q}) = \frac{2\pi \alpha}{|\vec{q}|}$ is the 2D Fourier transform of the Coulomb interaction and $\alpha$  is the effective fine-structure constant (we set the electronic charge $e=1$). Here, $\chi_0(\vec{q},\omega)$ is the non-interacting density-density correlation function (dynamical polarization function) of the system in response to an external perturbation, which, in the Fourier space reads as,
	\begin{align}
	\chi_0(\vec{q},\omega)=\frac{1}{A}\sum_{m,m^\prime,\vec{k}} F_{m,m^\prime}(\vec{k},\vec{q})~ \frac{ n_F(\xi_{m,\vec{k}}) - n_F(\xi_{m^{\prime},\vec{k}+\vec{q}}) }{\hbar \omega + i \eta + \xi_{m,\vec{k}} - \xi_{m^{\prime},\vec{k}+\vec{q}} }.\label{eq:chi0}
	\end{align}
Here $F_{m,m^\prime}(\vec{k},\vec{q})$ is the overlap between the eigenstates labeled by $\phi_{m,\vec{k}}$ and $\phi_{m^{\prime},\vec{k}+\vec{q}}$, given by
 \begin{align}
 F_{m,m^\prime}(\vec{k},\vec{q})=| \phi^{\dagger}_{m,\vec{k}}~ \phi_{m^{\prime},\vec{k}+\vec{q}} |^2 
 \end{align}	
At the zero temperature, the Fermi distribution function $n_F(\xi_{m,\vec{k}})$ turns into the simple step function $\theta(\mu-\xi_{m,\vec{k}})$, where $\mu$ is the Fermi energy. In rest of the calculation we will set $t=1$, which is our unit of energy. Other energy-scales, such as $\hbar\omega$ and $t_{so}$ are made unitless with respect to $t$, and we simply write $\hbar\omega/t$ as $\omega$.  We also consider a positive finite chemical potential $\mu$ for our numerical simulation.

The imaginary part of the non-interacting dynamical polarization function, $\chi_0(\vec{q},\omega)$, describes particle-hole (p-h) excitations, i.e., the process where an electron from an occupied state of wave vector $\vec{k}$ below Fermi energy is excited to an unoccupied one of wave vector $\vec{k}+\vec{q}$ above the Fermi energy leaving a hole below the Fermi energy. The region of the $(\vec{q},\omega)$ plane where ${\rm Im} \chi_0(\vec{q},\omega)$ is non-zero, is referred to as the particle-hole continuum (PHC).

If the Fermi energy is not far from the band touching points, we can neglect the particle-hole processes involving two momenta near two different Dirac points. These processes require large momentum transfer, and the contribution to the total polarization function is negligible. Neglecting these inter-nodal processes, we can write the total non-interacting dynamical polarization function as,
 \begin{align}
 \chi_0(\vec{q},\omega) \approx \chi_0^{X_1} + \chi_0^{X_2}+ \chi_0^{M} ,\label{eq:linchi}
 \end{align}
where $\chi_0^{X_1}$, $\chi_0^{X_2}$ and $\chi_0^{M}$ are the non-interacting dynamical polarization function near the Dirac points $X_1$, $X_2$ and $M$ respectively.

To proceed further, we expand the Hamiltonian up to linear order in $\vec{k}$ near each of the Dirac nodes; we find the linear-order Hamiltonians near the three Dirac nodes as,
\begin{equation}
\!
\begin{aligned}
	&H_{X_1}=-t_{so}\sigma_x k_y -t_{so}\tau_z\sigma_y k_x -t\sigma_z k_x ,\\
	&H_{X_2}=t_{so}\sigma_x k_y +t_{so}\tau_z\sigma_y k_x -t\sigma_z k_y,\\
	&	H_{M}=t_{so}\sigma_x k_y -t_{so}\tau_z \sigma_y k_x.
\end{aligned}\label{linH}
\end{equation}
The energy dispersion near the three Dirac nodes has the following structure,
	\begin{align}
   	\xi^{X_1}=\pm \sqrt{(t_{so}^2+t^2)k_x^2+t_{so}^2k_y^2},
	\end{align}
	
	\begin{align}
	\xi^{X_2}=\pm \sqrt{t_{so}^2k_x^2+(t_{so}^2+t^2)k_y^2},
	\end{align}
	and 
	\begin{align}
	\xi^{M}=\pm t_{so} k, ~~~~~ k=\sqrt{k_x^2+k_y^2}.
	\end{align} 
	Here each of the bands is doubly degenerate due to the presence of both inversion and time-reversal symmetry. The $\pm$ sign correspond to the positive and negative branch of energy dispersion.  Evidently, near Dirac nodes $X_1$ and $X_2$, the dispersion is anisotropic, but the dispersions at $X_1$ and $X_2$ are related by the symmetry $k_x \rightarrow k_y$. By considering the  linearized Hamiltonians, and the approximation Eq.~(\ref{eq:linchi}), we can write the total dynamical polarization function in the following form,
	\begin{align}
\chi_0(\vec{q},\omega)\approx \frac{1}{A}\sum_{P,\lambda,\lambda^\prime,\vec{k}} F^P_{\lambda,\lambda^\prime}(\vec{k},\vec{q})~ \frac{ n_F(\xi^P_{\lambda,\vec{k}}) - n_F(\xi^P_{\lambda^{\prime},\vec{k}+\vec{q}}) }{\hbar \omega + i \eta + \xi^P_{\lambda,\vec{k}} - \xi^P_{\lambda^{\prime},\vec{k}+\vec{q}} }.\label{eq:chi0l}
\end{align}
Here $P=X_1,X_2, M$ are the Dirac nodes, and $\lambda,\lambda'=\pm 1$ denote the positive and negative bands.

	\section{Results}\label{sec:result}
The plasmon dispersion and PHC for a finite, positive, chemical potential are shown in the right panel of Figure \ref{pl_dis}. We compute the plasmon mode numerically using the tight-binding Hamiltonian Eq.~(\ref{nonsym_tbm}), as well as using the linearized Hamiltonian Eq.~(\ref{linH}). The result obtained using the two models matches closely for small $q$, which is indeed expected, as for small $q$, the particle-hole excitations contributing to the collective plasmonic modes are mostly restricted to intra-node processes. 

\begin{figure}[t]
	\centering
		\includegraphics[width=0.4\textwidth]{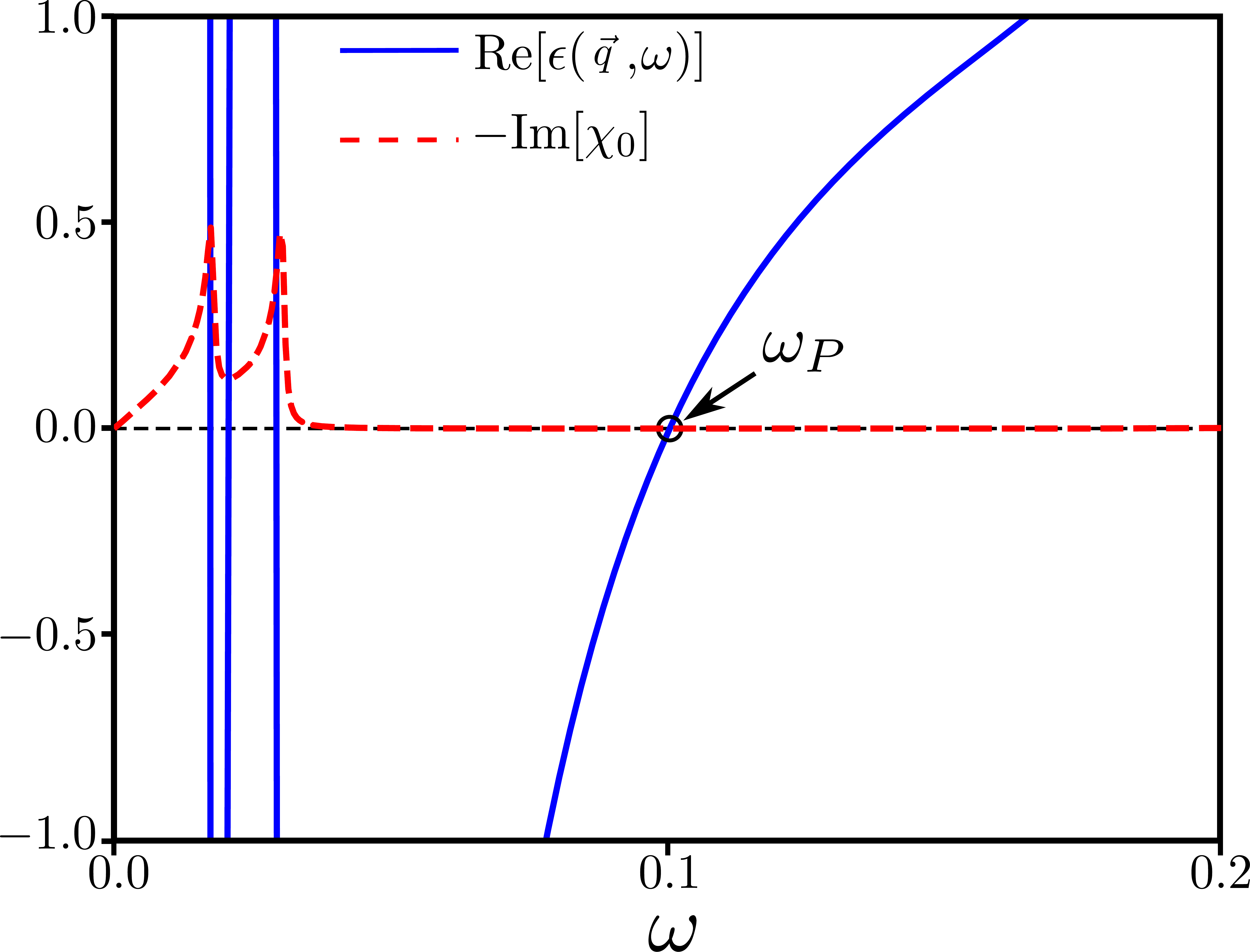}
		\caption{Plot of real part of dielectric function ${\rm Re}[\epsilon(\vec{q},\omega)]$(solid blue line) vs $\omega$ and ${\rm Im}[\chi_0(\vec{q},\omega)]$(red dashed line) vs $\omega$ for $q=0.035$. The ${\rm Re}[\epsilon(\vec{q},\omega)]$ vanishes at plasmon frequency $\omega_P$ as marked by a small circle. The imaginary part of $\chi_0$ also goes to zero at $\omega_P$.  }
		\label{rIchi0}
\end{figure}

 Plasmon modes can be observed experimentally in the electron energy loss spectrum, which directly measures the loss function, $-{\rm Im}[1/\epsilon(\vec{q},\omega)]$~\cite{lossf}. We show the density plot of loss function in the left panel of Fig. \ref{pl_dis}. The loss function is given by the following expression,
 \begin{align}
-{\rm Im}\Big[\frac{1}{\epsilon(\vec{q},\omega)}\Big] = \frac{V(|\vec{q}|) 
 	{\rm Im}[\chi^0]}{(1-V(|\vec{q}|){\rm Re}[\chi^0])^2+(V(|\vec{q}|){\rm Im}[\chi^0])^2}. \label{lossdefi}
\end{align}
A sharp (undamped) plasmon mode is obtained when ${\rm Re}[\epsilon(\vec{q},\omega)]=0$ and ${\rm Im}[\chi_0]=\eta$, a very small number, which immediately implies that loss function  is a $\delta$ function for undamped plasmon mode. The sharp, bright-line outside the PHC regime of the density plot of Fig. \ref{pl_dis} indicates the undamped plasmon mode. The plasmon mode gets damped significantly on entering the interband PHC.

We also show the plasmon modes in the $(\omega, \theta)$ plane for a fixed value of $\vec{q}$ in Fig.~\ref{iso}(c), where $\theta={\rm tan^{-1}} \frac{q_y}{q_x}$. To contrast the isotropic dispersions of the plasmon mode, we also consider the case of such collective modes in the presence of only one Dirac node at $X_1$ or $X_2$. Fig. \ref{iso}(a) and \ref{iso}(b) show the plasmon dispersions, if the system had only one such Dirac node, $X_1$ and $X_2$, respectively. Such plasmon dispersions are evidently anisotropic as a consequence of the anisotropies of these Dirac nodes. Whereas the whole system shows isotropic plasmon dispersion as  in the Fig. \ref{iso}(c). Fig.~\ref{rIchi0} shows the typical nature of the real part of $\epsilon(\vec{q},\omega)$ and the imaginary part of $\chi_0(\vec{q},\omega)$ as a function of $\omega$ a given momentum. At the plasmon frequency, the real part of $\epsilon(\vec{q},\omega)$ as well as the imaginary part of $\chi_0(\vec{q},\omega)$ vanishes.

\begin{figure}[t]
	\centering
	\includegraphics[width=0.4\textwidth]{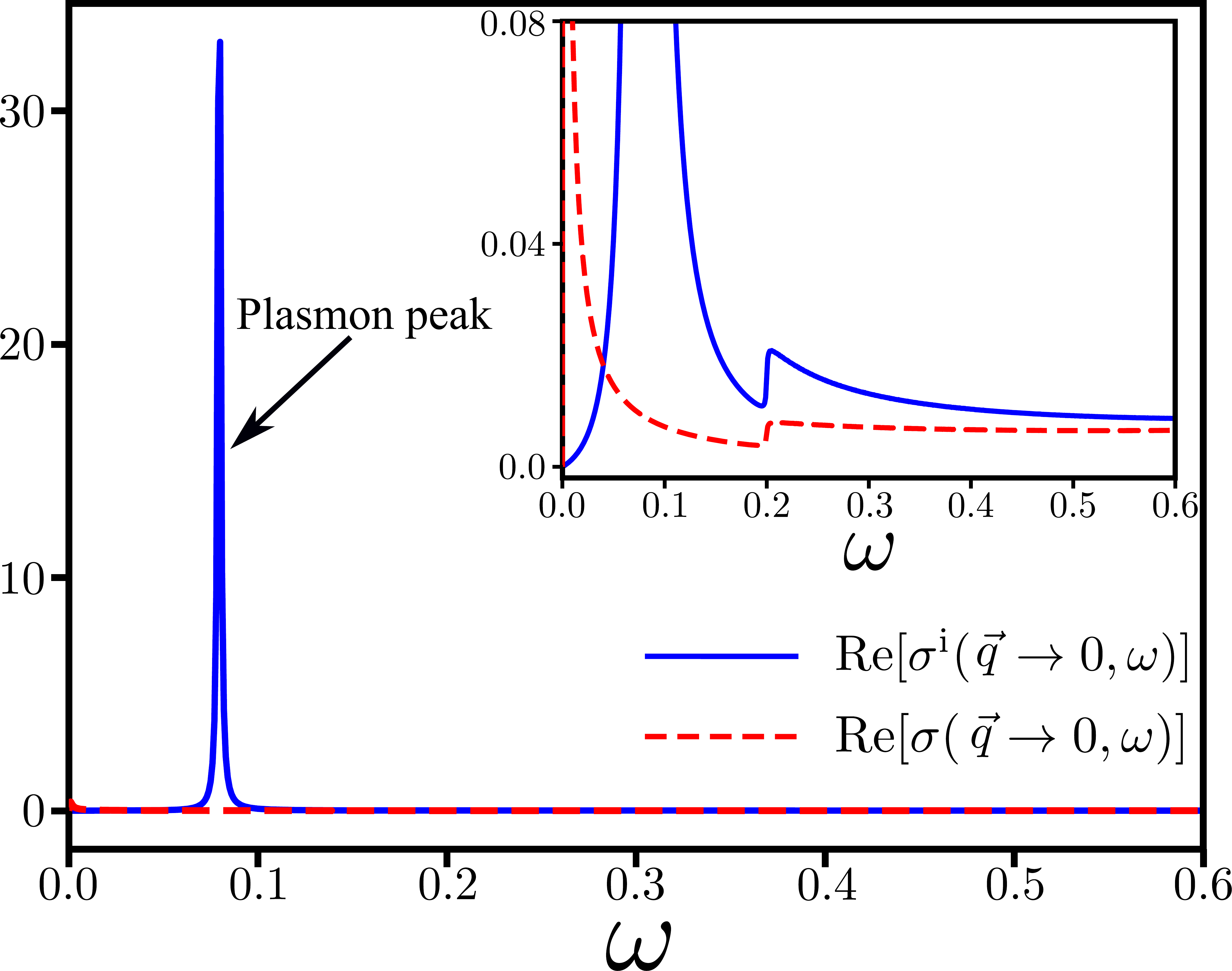}
	\caption{ Real part of optical conductivity
		$\sigma^{\rm{i}}(\vec{q}\to 0,\omega)$(blue solid line,interacting) and $\sigma(\vec{q}\to 0,\omega)$(red dashed line,non-interacting) vs $\omega$.The optical conductivity is in units of $e^2/\hbar$. Real part of optical conductivity  peaks up at plasmon frequency. }
	\label{optical}
\end{figure}

Plasmon modes also contribute to optical conductivity. The optical conductivity, $\sigma$, is related to the dynamical polarization functions in the non-interacting and interacting limit as following:
\begin{align}
	\sigma(\vec{q},\omega)=\frac{i\omega}{q^2}\chi_0(\vec{q},\omega)
\end{align}	
for the non-interacting system, whereas, 
\begin{align}
	\sigma^{\rm{i}}(\vec{q},\omega)=\frac{i\omega}{q^2}\chi(\vec{q},\omega),
\end{align}
for the interacting system. The real part of the optical conductivity corresponds to the dissipation of energy, which is given by,
\begin{align}
	{\rm Re}[\sigma^{\rm{i}}(\vec{q},\omega)]=-\frac{\omega}{q^2}{\rm Im}[\chi(\vec{q},\omega)].
\end{align}
We show in Fig.~\ref{optical} how the optical conductivity behaves as a function of frequency where the plasmon mode appears as an additional peak for the interacting system.

\begin{figure*}[htbp]
\begin{center}\leavevmode
 		\includegraphics[width=1.0\textwidth]{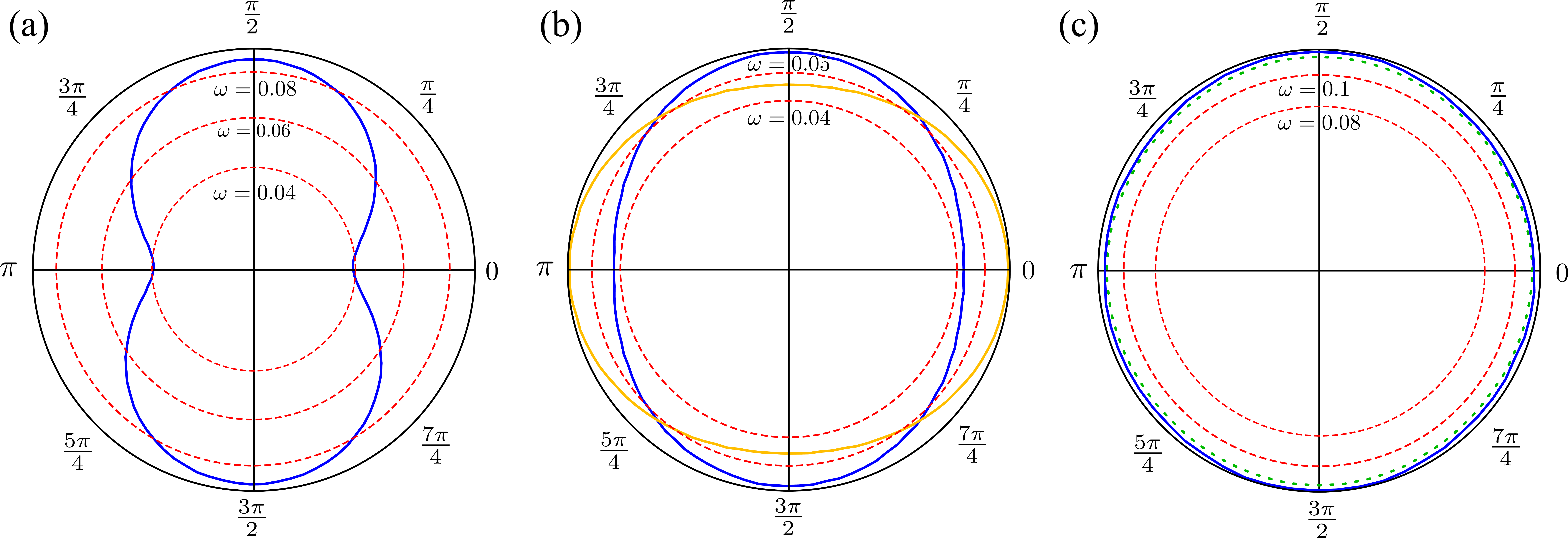}
 		\caption{ The solid curves represent the dispersion of the plasmon mode, for the system of Dirac nodes, Eq.~(\ref{linH}), along with the perturbations Eq.~(\ref{pert1}) and Eq.~(\ref{pert2}),  as a function of $(\omega,\theta)$ for a fixed value of $q=0.05$, where $\theta={\rm tan^{-1}} \frac{q_y}{q_x}$. The red-dashed circles are the locus of constant $\omega$. Panel (a): the plasmon dispersions, if one considers any of the split Weyl nodes at $X_2^{\pm}$ in the band-structure (see the main text). The dispersions clearly shows the anisotropic behavior of plasmon mode, i.e. $\omega$ changes as a function of $\theta$ and one can compare with Fig.~\ref{iso}(b). Panel  (b): In presence of the perturbation, near the $M$ point, the Hamiltonian is block-diagonal in two Weyl blocks, each of which are anisotropic. The plasmon dispersions shown in solid lines (blue and yellow), if one considers each of these Weyl blocks at the $M$ point in the band-structure, which are also anisotropic individually. The net effect of these two blocks, in the charge oscillation, is almost isotropic (not shown), for small $v_1$. Panel  (c): the plasmon dispersion of the nonsymmorphic Dirac semimetal which hosts two Dirac nodes, at momentum $X_1$, and $M$, and two Weyl nodes at momentum $X_2^{\pm}$ in the band-structure. The plasmon dispersion(blue solid curve) of this full system remains almost isotropic for the parameters we use, with a small anisotropy goes as $\sim v_1^2$. In comparison, the dotted (green) contour of the plasmon dispersion is obtained from the tight-binding Hamiltonian, Eq.~(\ref{nonsym_tbm}) along with the perturbations Eq.~(\ref{pert}). The parameters used are the same as in the Fig.~\ref{pl_dis} along with $v_1=0.2$. }
 		\label{pert}
\end{center}
\end{figure*}
	
	\subsection*{long wavelength limit($q\to 0$)}
Here we find the plasmon dispersion analytically in the long wavelength ($q\ll k_F$) and high frequency ($\hbar v_F q\ll \omega$) limit, where the $\chi_0$ can be expanded in the series of powers of $q/\omega$, yielding, for a generic system
\begin{align}
\chi_0(\vec{q},\omega) = r_1 \frac{q}{\omega} + r_2 \frac{q^2}{\omega^2} +\cdots,\label{eq:genchi}
\end{align}
where $r_i$s can be anisotropic. The first order term in $q$ vanishes in time-reversal symmetric systems. Keeping terms up to the second order in $q$, for time-reversal symmetric systems, the condition for the zeros of the dielectric function, Eq.~(\ref{eq:epsilon}), then gives,
\begin{align}
&1- \frac{2\pi \alpha}{q}\frac{r_2 q^2}{\omega^2} = 0 \Rightarrow~ \omega = \sqrt{2\pi\alpha r_2}\sqrt{q}.
\end{align}
For our system, in this regime of long-wavelength, $\chi_0(\vec{q},\omega)$ is mostly dominated by the p-h excitation processes near the Fermi energy, and one can neglect any inter-band processes even within the Dirac cones. We consider the chemical potential $\mu$ to be in the positive band, and, keeping only the intra-band processes, the non-interacting polarization function is written as,
 \begin{align}
\chi_0(\vec{q},\omega)=\frac{1}{A}\sum_{\vec{k}} F_{+,+}(\vec{k},\vec{q})~ \frac{ n_F(\xi_{+,\vec{k}}) - n_F(\xi_{+,\vec{k}+\vec{q}}) }{\omega  + \xi_{+,\vec{k}} - \xi_{+,\vec{k}+\vec{q}} }\label{eq:chi0s}
\end{align}
where $F_{+,+}(\vec{k},\vec{q})=| \phi^{\dagger}_{+,\vec{k}}~ \phi_{+,\vec{k}+\vec{q}} |^2 $, with $\phi_{+}$ are the eigenstates corresponding to the positive branch of the energy. 

Let us consider the case near one of the Dirac point, say at $X_1$. In the long wavelength and high frequency limit, 
\begin{align}
&\xi_{\vec{k}+\vec{q}}^{X_1}-\xi_{\vec{k}}^{X_1}\approx \frac{(t^2+t^2_{so})q_xk_x}{\xi^{X_1}_{\vec{k}}} +\frac{t^2_{so}q_yk_y}{\xi^{X_1}_{\vec{k}}},
\end{align}
and $| \phi^{\dagger}_{+,\vec{k}}~ \phi_{+,\vec{k}+\vec{q}} |^2\approx 1 + \mathcal{O}(q^2)$, where we drop the $\mathcal{O}(q^2)$ term as we wish to expand $\chi_0(\vec{q},\omega)$ up to order $q^2$, as following. We write
 \begin{align}
\chi_0(\vec{q},\omega)^{X_1}&=\frac{1}{A}\sum_{\vec{k}} ~ \frac{ n_F(\xi^{X_1}_{+,\vec{k}}) - n_F(\xi^{X_1}_{+,\vec{k}+\vec{q}}) }{\omega  + \xi^{X_1}_{+,\vec{k}} - \xi^{X_1}_{+,\vec{k}+\vec{q}} }\nonumber \\
&\approx-\frac{1}{4\pi^2}\int d^2\vec{k}~ \frac{\partial n_F(\xi^{X_1}_{+,\vec{k}})}{\partial \xi^{X_1}_{+,\vec{k}}}~  ~\frac{  \xi^{X_1}_{+,\vec{k}+\vec{q}}-\xi^{X_1}_{+,\vec{k}}}{\omega  + \xi^{X_1}_{+,\vec{k}} - \xi^{X_1}_{+,\vec{k}+\vec{q}}  }\nonumber \\ 
&\approx\frac{1}{4\pi^2}\int d^2\vec{k}~~ \delta(\xi^{X_1}_{+,\vec{k}}-\mu)~\frac{\xi^{X_1}_{+,\vec{k}+\vec{q}}-\xi^{X_1}_{+,\vec{k}}}{\omega }\nonumber\\
&\quad\quad\quad\quad\quad\quad \times \bigg(1-\frac{\xi^{X_1}_{+,\vec{k}} - \xi^{X_1}_{+,\vec{k}+\vec{q}}}{\omega }\bigg).\label{eq:chi0appx}
\end{align}
Let $\sqrt{t^2+t_{so}^2}~k_x=w\cos{\theta}$ and $t_{so} ~ k_y=w\sin{\theta}$; the Jacobian for such coordinate transformation is $dk_x dk_y=\frac{1}{\sqrt{t^2+t_{so}^2} ~t_{so}} wdw d\theta$.
Therefore,
\begin{align}
\chi_0(\vec{q},\omega)^{X_1}&=\frac{1}{4\pi^2\omega \sqrt{t^2+t_{so}^2}~ t_{so}} \int w dw~ \delta (w-\mu)\int_{0}^{2\pi}d\theta\nonumber\\
&\bigg[ \big(\sqrt{t^2+t_{so}^2}~ q_x \cos{\theta}+t_{so}~q_y\sin{\theta}\big)\nonumber\\
& + \frac{\big(\sqrt{t^2+t_{so}^2}~ q_x \cos{\theta}+t_{so}~q_y\sin{\theta}\big)^2}{\omega} \bigg].
\end{align}
The first term vanishes after the integration over $\theta$ and the second term gives 
\begin{align}
	\chi_0(\vec{q},\omega)^{X_1}&=\frac{\mu}{4\pi \sqrt{t^2+t_{so}^2}~ t_{so}}\frac{(t^2+t_{so}^2)q_x^2 +t_{so}^2 q_y^2}{\omega^2}.
\end{align}
Near the other Dirac nodes, $\chi_0(\vec{q},\omega)^{X_2}$ and $\chi_0(\vec{q},\omega)^{M}$ can be calculated following the same procedure, yielding,
\begin{align}
&	\chi_0(\vec{q},\omega)^{X_2}=\frac{\mu}{4\pi\sqrt{t^2+t_{so}^2} ~t_{so} } \frac{~[ t_{so}^2 q_x^2+(t^2+t_{so}^2)q_y^2 ]}{\omega^2},\label{eq:chi0X2}\\
&	\chi_0(\vec{q},\omega)^{M}=\frac{\mu}{4\pi } \frac{q^2}{\omega^2}.\label{eq:chi0M}
\end{align}
Evidently, the net $\chi$ becomes isotropic. Using these in the Eq. (\ref{eq:epsilon}) we get,
\begin{align}
 \omega=\sqrt{\mu \alpha\left(\frac{t^2+2t_{so}^2}{\sqrt{t^2+t_{so}^2} ~t_{so}} +1 \right)}\sqrt{q}.\label{eq:omega}
\end{align}
which implies an isotropic dispersion of the single plasmon mode.

In passing, we also comment that if we had included the next nearest term $t_2$, in the Hamiltonian, Eq.~(\ref{nonsym_tbm0}), that would result in a shift in the energy for the Dirac node at the $M$ point. As the anisotropic Dirac nodes at $X_1$ and $X_2$ remain at the same energy, our analysis would broadly follow where the dominant contributions in $\chi_0$ appearing from only $\chi_0^{X_1}$ and $\chi_0^{X_2}$, still resulting in an isotropic plasmon dispersion.

One can further calculate the dependency of the plasmon frequency on the carrier density ($n$) as follows. The density of states near the Dirac point $X1$, $X2$ and $M$ are given by
\begin{align}
&{D(E)_{X_1}=D(E)_{X_2}=\frac{E}{\pi t_{so}\sqrt{t^2+t_{so}^2}}}, \nonumber \\
{\rm and}~~ &{D(E)_{M}=\frac{E}{\pi t_{so}^2}}.
\end{align}
Thus, the net carrier density $n$ is
\begin{align}
n=\int_{0}^{\mu} D(E) dE =\frac{\mu^2}{2\pi}\frac{2t_{so}+\sqrt{t^2+t_{so}^2}}{t_{so}^2 \sqrt{t^2+t_{so}^2}}. \label{eq:n}
\end{align}
Comparing Eq.~(\ref{eq:omega}) and Eq.~(\ref{eq:n}), it is easy to show that $\omega \propto n^{1/4}q^{1/2}$ as in the case of pristine Graphene \cite{density}.

\section{Discussion}\label{sec:sum}
Even though we assume that the Dirac nodes at $X_1$ and $X_2$ are related by symmetries, which is responsible for the isotropic dispersion of the plasmon mode, such symmetry may not be present for a realistic materials~\cite{bism}. In such cases, the parameters $t$ and $t_{so}$ can be different for these two nodes, and the resultant plasmon mode can become anisotropic. If we write
\begin{align}
&	\chi_0(\vec{q},\omega)^{X_1}=\frac{\mu}{4\pi \sqrt{t^2+t_{so}^2}~ t_{so}}\frac{(t^2+t_{so}^2)q_x^2 +t_{so}^2 q_y^2}{\omega^2}\nonumber\\
&	\chi_0(\vec{q},\omega)^{X_2}=\frac{\mu}{4\pi\sqrt{t^{\prime2}+t_{so}^{\prime2}} ~t_{so}^{\prime} } \frac{~[ t_{so}^{\prime2} q_x^2+(t^{\prime2}+t_{so}^{\prime2})q_y^2 ]}{\omega^2},
\end{align}
and assuming, for simplicity, in the absence of the Dirac node at $M$ at the same energy, we obtain, (keeping the notational similarity with Eq.~(\ref{eq:genchi})),
\begin{align}
\chi_0(\vec{q},\omega) = r_2 \frac{q^2}{\omega^2},
\end{align}
with
\begin{align}
r_2 =& \frac{\mu}{4\pi}\left(\frac{(t^2+t_{so}^2)\cos^2\theta +t_{so}^2 \sin^2\theta}{\sqrt{t^2+t_{so}^2}~ t_{so}}\right.\nonumber\\
& \quad\quad + \left. \frac{(t^{2\prime}+t_{so}^{\prime2})\sin^2\theta +t_{so}^{\prime2} \cos^2\theta}{\sqrt{t^{\prime2}+t_{so}^{\prime2}}~ t_{so}^{\prime}}\right),
\end{align}
resulting in an anisotropic plasmon mode with dispersion $\omega = \sqrt{2\pi\alpha r_2}\sqrt{q}$. The degree of the anisotropy is given by the ratio
\begin{align}\label{eq:anisdeg}
\mathcal{A} = \frac{t^2 + (t_{so}^2+ t_{so}^{\prime2})}{t^{\prime2} + (t_{so}^2+ t_{so}^{\prime2})}.
\end{align}
Thus, detecting the plasmon mode can provide further information of the details of the underlying band-structure.

Additionally, we also consider a perturbation of the form (written in the rotated basis, given by Eq.~(\ref{eq:U})),
\begin{align}\label{eq:pert}
V = -\sigma_y v_1 \sin \frac{k_y}{2}\cos \frac{k_x}{2}.
\end{align}
This term breaks the time-reversal symmetry, but keeps the $C_{2x}$ and $C_{2y}$ symmetries intact. The effect of this term is to split the Dirac node at the $X_2$ point into a pair of Weyl nodes, at position $X_2^{\pm}=(\pm 2\sin^{-1}(v_1/2t_{so}),\pi)$. Although it keeps the low-energy Hamiltonian at $X_1$ intact, this perturbation also makes the Dirac node at $M$ point anisotropic. The low-energy Hamiltonians at the two Weyl nodes at $X_2^{\pm}$ reads, up to order $v_1^2$ and linear in $k$, as:
\begin{align}\label{pert1}
H^{\pm}_{X_2} =t_{so}\sigma_x k_y &\pm t_{so}\sigma_y k_x -t\sigma_z k_y\nonumber\\
& \mp \frac{v_1^2}{4t_{so}}\sigma_y k_x  +\frac{tv_1^2}{8t_{so}^2} \sigma_z k_y,
\end{align}
where each of $H^{\pm}_{X_2} $ are of dimension two, instead of four. In addition, it introduces a perturbation in the $H_M$ as
\begin{align}\label{pert2}
\Delta H_M = \frac12 v_1\sigma_y k_x.
\end{align}
The effect of this perturbation in the band structure is also shown in the Fig.~\ref{bands}. Even though the band-touching is four-fold degenerate at the $M$ point, the positive (as well as negative) bands near the $M$ point are no anymore degenerate. Essentially, the Hamiltonian near the $M$ point is now block-diagonal in two Weyl cones, each of which is anisotropic. The effect of this perturbation in the plasmon dispersion, for the small $q$ regime, has been plotted in Fig.~\ref{pert}. The perturbation introduces anisotropy to each block at the Hamiltonian near the $M$ point. Although the dispersions of  each of the $H_{\chi_2}^{\pm}$ is also anisotropic, the overall charge oscillation taken into account all the Dirac and Weyl nodes is very weakly anisotropic, as shown in Fig.~\ref{pert}(c) for the parameters we use. The anisotropy grows as $v_1^2$.


In summary, we studied the plasmon dispersion and the
 loss-function for the nonsymmorphic Dirac semimetal introduced in the Hamiltonian Eq.~(\ref{nonsym_tbm0}) that contains three Dirac nodes, two of which are anisotropic in their dispersion. We obtain a single, isotropic plasmon mode originating from the collective charge oscillations in the system in the presence of electron-electron interaction. Such isotropic dispersion is an outcome of the symmetry between the two anisotropic Dirac nodes. In the absence of such symmetry, the plasmon mode can be anisotropic with a degree of anisotropy given by Eq.~(\ref{eq:anisdeg}). We also consider the effect of a time-reversal breaking perturbation, that splits one of the Dirac nodes into two Weyl nodes, in the dispersion of the resulting plasmon mode. As far as experimental observations are concerned, $\alpha$-Bi resembles a similar band structure of the nonsymmorphic Dirac system, although in this compound, the Dirac nodes are known to be somewhat away from the Fermi energy~\cite{bism}. Nevertheless, our predictions of the plasmon dispersion should be valid in similar nonsymmorphic systems. The detection of such plasmon modes through electron energy-loss spectroscopy may reveal the nature of the electronic bands of these systems. Finally, similar to Graphene, we find the plasmon frequency of the nonsymmorphic Dirac systems is also tunable, such as by changing the carrier concentration through external doping, allowing them to be possible candidates for terahertz technologies.

\section{Acknowledgments}
A.K  acknowledges support from the SERB (Govt. of India) via saction no. ECR/2018/001443, DAE (Govt. of India ) via sanction no. 58/20/15/2019-BRNS, as well as MHRD (Govt. of India) via sanction no. SPARC/2018-2019/P538/SL. D.G acknowledges the CSIR (Govt. of India) for financial support. We also  acknowledge the use of HPC facility at IIT Kanpur.

	\appendix{}
	\begin{widetext}
	\section{density-density correlation}\label{2nd_not}
	Using the definitions of the density and the field operators one can find,
	\begin{align}\nonumber
	[\rho(\vec{q},t) , \rho(-\vec{q})]&=\sum \phi^{\dagger}_{m,\vec{k}}  ~\phi_{m^{\prime},\vec{k}+\vec{q}} ~\phi^{\dagger}_{s,\vec{k}^{\prime}} ~\phi_{s^{\prime},\vec{k}^{\prime}-\vec{q}}~\big\langle \big[ \exp(i  H t)~c^{\dagger}_{m,\vec{k}}~ c_{m^{\prime},\vec{k}+\vec{q}} ~\exp(-i  H t)~, ~ c^{\dagger}_{s,\vec{k}^{\prime}} ~c_{s^{\prime},\vec{k}^{\prime}-\vec{q} }\big]\big \rangle.
	\end{align}
	Therefore,
	\begin{align}
	\chi(\vec{q},t)&=\frac{1}{A}\sum \chi_{m,m^{\prime},s,s^{\prime}}(\vec{k},\vec{k}^{\prime},\vec{q},t)
	\end{align}
	where,
	\begin{align}\nonumber
	\chi_{m,m^{\prime},s,s^{\prime}}(\vec{k},\vec{k}^{\prime},\vec{q},t)&=-i \Theta(t) \phi^{\dagger}_{m,\vec{k}}  ~\phi_{m^{\prime},\vec{k}+\vec{q}} ~\phi^{\dagger}_{s,\vec{k}^{\prime}} ~\phi_{s^{\prime},\vec{k}^{\prime}-\vec{q}}~\\
	&\big\langle \big[ \exp(i  H t)~c^{\dagger}_{m,\vec{k}}~ c_{m^{\prime},\vec{k}+\vec{q}} ~\exp(-i  H t)~, ~ c^{\dagger}_{s,\vec{k}^{\prime}} ~c_{s^{\prime},\vec{k}^{\prime}-\vec{q} }\big]\big \rangle \label{eq:chif}
	\end{align}
	One can find the equation of motion of Eq. \ref{eq:chif}, which is given by,
	\begin{align}
	\partial_t ~\chi_{m,m^{\prime},s,s^{\prime}}(\vec{k},\vec{k}^{\prime},\vec{q},t)&=-i\delta(t)~\phi^{\dagger}_{m,\vec{k}}~ \phi_{m^{\prime},\vec{k}+\vec{q}}  ~\phi^{\dagger}_{s,\vec{k}^{\prime}}  ~\phi_{s^{\prime},\vec{k}^{\prime}-\vec{q}}~ \big\langle \big[~c^{\dagger}_{m,\vec{k}^{\prime}}~ c_{m^{\prime},\vec{k}+\vec{q}}~,  ~c^{\dagger}_{s,\vec{k}^{\prime}} ~c_{s^{\prime},\vec{k}^{\prime}-\vec{q} } ~\big]\big \rangle \nonumber\\
	&-i\Theta(t) ~\phi^{\dagger}_{m,\vec{k}} ~ \phi_{m^{\prime},\vec{k}+\vec{q}}  ~\phi^{\dagger}_{s,\vec{k}^{\prime}} ~\phi_{s^{\prime},\vec{k}^{\prime}-\vec{q}} 
		~\big\langle i \big[ \exp(i  H t)~[H~,~c^{\dagger}_{m,\vec{k}^{\prime}}~ c_{m^{\prime},\vec{k}+\vec{q}}]~ \exp(-i  H t),  ~c^{\dagger}_{s,\vec{k}^{\prime}} ~c_{s^{\prime},\vec{k}^{\prime}-\vec{q} }~\big]\big \rangle\label{eq:partial}
	\end{align}
	By using anti-commutation rule of fermionic operator, we write,
	\begin{align}
&\big\langle	\big [~c^{\dagger}_{m,\vec{k}}c_{m^{\prime},\vec{k}+\vec{q}}     ~   ,~ c^{\dagger}_{s,\vec{k}^{\prime}}c_{s^{\prime},\vec{k}^{\prime}-\vec{q}}    ~    \big]\big \rangle~=~ \big[~n_F\left(\xi_{m,\vec{k}}\right) - n_F\left(\xi_{m^{\prime},\vec{k}+\vec{q}}\right)~\big]~ \delta_{m,s^{\prime}}~\delta_{m^{\prime},s}~ \delta_{\vec{k}+\vec{q},\vec{k}^{\prime}},\label{eq:t1}\\
&	\big[~ H_0~,~ c^{\dagger}_{m,\vec{k}}c_{m^{\prime},\vec{k}+\vec{q}}~\big]= \left(\xi_{m,\vec{k}} -\xi_{m^{\prime},\vec{k}+\vec{q}} \right)~c^{\dagger}_{m,\vec{k}}c_{m^{\prime},\vec{k}+\vec{q}}~.\label{eq:t2}
	\end{align}

\subsubsection*{Hartree approximation}
The term 
$
\big[~ H_{int}~,~ c^{\dagger}_{m,\vec{k}}c_{m^{\prime},\vec{k}+\vec{q}}~\big]
$ generates two-particle operators. One applies the Hartree approximation to reduce the into single particle operators by replacing $c^\dagger_{\vec{k}} c_{\vec{k}^\prime}$ with its thermal average value, $\langle c^\dagger_{\vec{k}} c_{\vec{k}^\prime}\rangle$ .
Now,
 	\begin{align}
 	\big[~ H_{int}~,~ c^{\dagger}_{m,\vec{k}}c_{m^{\prime},\vec{k}+\vec{q}}~\big]~&=~\sum(\cdots)\bigg(~c^\dagger_{l_1,\vec{k}_1}c^\dagger_{l_2,\vec{k}_2}c_{l_3,\vec{k}_2+\vec{q}_1}c_{m^{\prime},\vec{k}+\vec{q}} ~\delta_{l_4,m}~\delta_{\vec{k}_1-\vec{q}_1,\vec{k}}\nonumber \\
 	&+ ~c^\dagger_{l_1,\vec{k}_1}c^\dagger_{l_2,\vec{k}_2} c_{m^{\prime},\vec{k}+\vec{q}}c_{l_4,\vec{k}_1-\vec{q}_1}~\delta_{l_3,m}~\delta_{\vec{k}_2+\vec{q}_1,\vec{k}} -~c^\dagger_{l_1,\vec{k}_1}c^{\dagger}_{m,\vec{k}}c_{l_3,\vec{k}_2+\vec{q}_1}c_{l_4,\vec{k}_1-\vec{q}_1}~\delta_{l_2,m^{\prime}}~\delta_{\vec{k}_2,\vec{k}+\vec{q}}                                            \nonumber\\
 	&-~c^{\dagger}_{m,\vec{k}}c^\dagger_{l_2,\vec{k}_2}c_{l_3,\vec{k}_2+\vec{q}_1}c_{l_4,\vec{k}_1-\vec{q}_1}~\delta_{l_1,m^{\prime}}~\delta_{\vec{k}_1,\vec{k}+\vec{q}}\bigg), \label{eq:int}
 	\end{align}
 where we do not explicitly write the terms multiplying the operators in $(\cdots) = {1 \over A} ~ V(\textbf{q}_1) \phi^\dagger_{l_1,\textbf{k}_1}~\phi_{l_4,\textbf{k}_1-\textbf{q}_1}~\phi^\dagger_{l_2,\textbf{k}_2}~\phi_{l_3,\textbf{k}_2+\textbf{q}_1}$. We replace the operator of the form $\langle c^\dagger_{\vec{k}_1} c_{\vec{k}_2+\vec{q}_j} \rangle$ by $n_F(\xi_{\vec{k}_1})\delta_{\vec{k}_1,\vec{k}_2+\vec{q}_j}$.
 	After some algebric calculations Eq.~(\ref{eq:int}) can be simplified as,
 	\begin{align}
 		\big[H_{int}~,~c^{\dagger}_{m,\vec{k}} c_{m^{\prime},\vec{k}+\vec{q}}\big]~=\sum(\cdots)\bigg(~2\big[ n_F(\xi_{m^{\prime},\vec{k}+\vec{q}})- n_F(\xi_{m,\vec{k}})\big] ~c^\dagger_{l_2,\vec{k}_2}c_{l_3,\vec{k}_2+\vec{q}_1}
 	~\delta_{l_4,m}~\delta_{l_1,m^{\prime}} ~\delta_{\vec{k}_1-\vec{q}_1,\vec{k}}~\delta_{\vec{k}_1,\vec{k}+\vec{q}}\bigg). \label{eq:HAf1}
 	\end{align}
Using Eq.~(\ref{eq:t1}), (\ref{eq:t2}) and (\ref{eq:HAf1}) in Eq.~(\ref{eq:partial}), we get,
 	\begin{align}
 		\partial_t ~\chi_{m,m^{\prime},s,s^{\prime}}(\vec{k},\vec{k}^{\prime},\vec{q},t)&=-i\delta(t) F_{m,m^\prime,}(\vec{k},\vec{q}) \big[~n_F(\xi_{m,\vec{k}}) - n_F(\xi_{m^{\prime},\vec{k}+\vec{q}})~\big]~ \delta_{m,s^{\prime}}~\delta_{m^{\prime},s}~ \delta_{\vec{k}+\vec{q},\vec{k}^{\prime}} \nonumber \\
 		&+i~(\xi_{m,\vec{k}}-\xi_{m^\prime,\vec{k}+\vec{q}})\chi_{m,m^\prime,s,s^\prime}(\vec{k},\vec{k}^\prime,\vec{q},t) \nonumber\\
 		&+i\big[n_F(\xi_{m^{\prime},\vec{k}+\vec{q}})-n_F(\xi_{m,\vec{k}})\big] V(\vec{q})F_{m,m^\prime,}(\vec{k},\vec{q})\frac1A\sum_{l_2,l_3,\vec{k}_2}\chi_{l_2,l_3,s,s^\prime}(\vec{k}_2,\vec{k}^\prime,\vec{q},t). \label{eq:HA2}
 	\end{align}
 Now we perform the Fourier transform of the Eq.~(\ref{eq:HA2}) from time, $t$, to frequency, $\omega$, and, then sum over the indices, $m$, $m^\prime$, $s$, $s^\prime$, $\vec{k}$, $\vec{k}^\prime$, to obtain,
 	\begin{align}
 	\chi(\vec{q},\omega)=\frac{\chi_0(\vec{q},\omega)}{1-V(\vec{q})\chi_0(\vec{q},\omega)}\label{eq:chifinal}
 	\end{align}
 with
  $\chi_0(\vec{q},\omega)$ being the non-interacting density-density response function given by Eq.~(\ref{eq:chi0}).

\end{widetext}


\begin{thebibliography}{99}
\bibitem{SDP}
Dirac points which are robust against spin-orbit coupling.
\end{thebibliography}

\end{document}